\documentclass[prl,twocolumn,showpacs]{revtex4}
\usepackage{graphicx}

\newcommand{\da}{\dagger}  
\newcommand{\be}{\begin{equation}}
\newcommand{\eq}{\end{equation}}
   
\newcommand{\Tr}{{\rm \, Tr \!}}    

\begin{document}

%
\title{Finite temperature  gauge theory from the
transverse lattice}
\author{S. Dalley}
\affiliation{Department of Physics, University of Wales Swansea, 
Singleton Park,
Swansea SA2 8PP, United Kingdom}
\author{B. van de Sande}
\affiliation{Geneva College,
3200 College Ave., Beaver Falls, PA 15010}
\preprint{SWAT/409}

%
%
%


\begin{abstract}
Numerical computations are performed and analytic
bounds are obtained on  the excited spectrum of glueballs 
in $SU(\infty)$ gauge theory, by transverse lattice Hamiltonian
methods. We find an 
exponential growth of the density of states, implying
a finite critical (Hagedorn) temperature. 
It is argued that 
the Nambu-Goto string model lies in a different universality class.
\end{abstract}

\maketitle



\section{Introduction}
\label{intro}

The behavior of QCD at finite temperature is of great
theoretical and experimental interest.  Traditionally, finite
temperature calculations of a field theory are performed 
in Euclidean space with periodic boundary conditions imposed 
in the time direction.  However, the same information can be
extracted from a Minkowski space calculation by examining
the density of states in the spectrum
as a function of energy.
In particular, an exponentially increasing density of 
states implies a finite critical temperature $T_c$~\cite{hagedorn}.
There is some experimental evidence of this exponential increase
in the hadron spectrum
\cite{bron}. 
Recently, there has been renewed interest
in the lightcone formulation of thermal field theory in Minkowski
space \cite{hot}. In this letter we investigate the aspects of
a lightcone formulation
of QCD relevant for finite temperature and, in particular, its
implications for the string picture of QCD.

To leading order of the $1/N_c$ expansion of QCD \cite{hoof},
we compute the density of glueball states per unit mass 
via the transverse lattice approach to non-abelian gauge 
theory~\cite{bard1}. This is
a Hamiltonian method that combines light-front quantization
in the two ``longitudinal'' spacetime directions, 
$x^\pm=(x^0\pm x^3)/\sqrt{2}$, with a lattice
in the remaining ``transverse'' directions, $\{x^1, x^2\}$.
It has been successfully used to compute the properties of the lightest
glueballs and mesons (for a review, see Ref.~\cite{trans}). 
In these calculations, effective Hamiltonians on coarse lattices  
were tuned to optimize Lorentz covariance of the lightest boundstates.
The method also yields the spectrum of heavier boundstates.
A rapid exponential rise in the density of states suggests that there may
be enough states at relatively
low masses,
where the coarse lattice calculation is still accurate 
for aggregate quantities,
to observe this behaviour unambiguously.

Using the detailed methods described in Refs.~\cite{dv3}, 
we demonstrate that in an intermediate mass range, $\sim 1-3$ times the
lightest glueball mass, the density of states does indeed
show a degree of universality with respect to cutoffs.
We find an exponentially increasing density of states, implying a 
finite-temperature Hagedorn transition.
Although the method is not well-suited to producing an accurate
prediction of the critical temperature, we make a numerical 
estimate and produce analytic bounds on it. One application
of interest is to the string picture of large-$N_c$ gauge theory.
Results from previous Euclidean lattice computations \cite{teper}
indicate a  $T_c$ lower than that of the Nambu-Goto string model
\cite{nambu},
suggesting  that a QCD string in the large-$N_c$ limit has more worldsheet
degrees of freedom than this free bosonic string.
Our calculations suggest that this is due at least in part to
longitudinal oscillations. 
This is also consistent
with the Nambu-Goto string description appearing in a distinct, unstable phase
of transverse lattice gauge theory discovered by Klebanov and Susskind 
\cite{kleb}.

\section{Transverse lattice gauge theory}

The transverse lattice is formulated in the following
manner.
Two coordinates $x^\alpha$, $\alpha \in \{0,3\}$, are continuous
while two directions ${\bf x}=\{x^1,x^2\}$ are discretized
as a lattice, spacing $a$.
The  longitudinal continuum
gauge potentials $A^\alpha({\bf x})$ 
lie at sites ${\bf x}$.
The transverse flux link fields $M_r({\bf x})$, $r\in\{1,2\}$, 
lie on the link between ${\bf x}$
and ${\bf x} + a \hat{\bf r}$,  representing the gauge
fields polarized in the ${\bf x}$ directions. In general
there are also fermi fields $\Psi({\bf x})$, but we will not need these
for the glueball spectrum at large-$N_c$.
For a coarse transverse lattice, 
the strategy is to perform
a color-dieletric expansion \cite{dv2} of the
most general lightcone gauge Hamiltonian, 
renormalisable with respect to
the continuum coordinates $x^\alpha$, in powers of $M_r$. 
Provided these link fields
are chosen sufficiently heavy, one can truncate this expansion to study the
low lying hadron boundstates dominated by just a few particles
of these fields. The remaining couplings in the effective
Hamiltonian can then 
be accurately constrained by optimizing symmetries broken by
the cutoffs in low energy observables. Typically this method is
viable in a  window
between small lattice spacings, where the fields become too light to
justify the color-dieletric expansion, and large lattice spacings where
Lorentz covariance breaks down uncontrollably.

For completeness, we give the Lagrangian, used in Ref.~\cite{dv3}, 
that we shall employ here.
It contains all allowed terms up to order  $(M)^4$ 
for the large-$N_c$ limit:
\begin{eqnarray}
L & = &  \sum_{{\bf x}} \int dx^- \sum_{\alpha, \beta = +,-}
\sum_{r=1,2} 
 \nonumber
\\
&& -{1 \over 2 G^2} \Tr \left\{ F^{\alpha \beta}
F_{\alpha \beta}
\right\}
- \mu_{b}^2  \Tr\left\{M_r M_r^{\da}\right\}
\nonumber \\
&& + \Tr\left\{[\left(\partial_{\alpha} + {\rm i} A_{\alpha} 
\right)
        M_r
-  {\rm i} M_r
A_{\alpha}
][\mbox{ H. C. }]\right\}
\nonumber \\
&& 
+{\beta \over N_c a^{2}} \Tr\left\{ M_{1} 
M_{2} 
M_{1}^{\da}
M_{2}^{\da}
\right\} + \mbox{H. C. } \nonumber \\
&& - \frac{1}{a^{2}} \sum_r\left[ \frac{\lambda_1}{N_c}
\Tr\left\{ M_r M_r^{\da}
M_r M_r^{\da} \right\} \right.
\nonumber \\ &&
\left. + \frac{\lambda_2}{N_c} 
\Tr\left\{ M_r 
M_r
M_r^{\da}
M_r^{\da} 
\right\} 
 + 
\frac{\lambda_3}{ N_{c}^2} 
\left( \Tr\left\{ M_r M_r^{\da} \right\} \right)^2\right]
\nonumber\\ &&
-  {\lambda_4 \over a^{2} N_c}  
\sum_{\sigma=\pm 2, \sigma^\prime = \pm 1}
        \Tr\left\{ 
M_\sigma^{\da} M_\sigma M_{\sigma^\prime}^{\da} M_{\sigma^\prime} \right\} 
        \nonumber\\
&& -  {4 \lambda_5 \over a^{2} N_{c}^2} 
\Tr\left\{ M_1 M_1^{\da} \right\}\Tr\left\{ M_2 M_2^{\da} \right\} 
\; . \label{pot1} 
\label{ferlag}
\end{eqnarray}
where $F^{\alpha \beta}({\bf x})$ is the continuum field strength in the
$(x^0,x^3)$ planes at each site ${\bf x}$.
We have defined $M_{r} = M_{-r}^{\dagger}$ and hold
$G^2 N_c$ finite as $N_c \to \infty$.
In the $N_c \to \infty$ limit, glueball states consists of connected loops
of transverse flux links and are absolutely
stable to decay.

The reader is referred to Ref.~\cite{dv3}
 for details of the 
construction of the lightcone Hamiltonian and Fock space, the renormalisation,
and the determination of various couplings appearing in 
Eqn.~(\ref{ferlag}). 
These couplings are accurately constrained by 
optimizing covariance of low-lying glueball eigenfunctions and
rotational invariance of the heavy-source potential. 
The relatively large number of couplings means the Hamiltonian is highly
`improved' and can give cut-off independent results on quite coarse lattices. 
Taking the  fundamental scale to be the string tension $\sigma$, 
Refs.~\cite{dv3} investigated individual glueball masses up to 
about 1.5 times the lightest glueball mass and
the method was shown to produce accurate results,
even for lattice spacings of order $1/\sqrt{\sigma}$.
We  investigate here whether any part of 
the higher excited spectrum, probed at finite temperature, is similarly
under control. 

\begin{figure*}
$\displaystyle\ln (t)$\hspace{5pt}
\raisebox{-1.75in}{\includegraphics[width=5in]{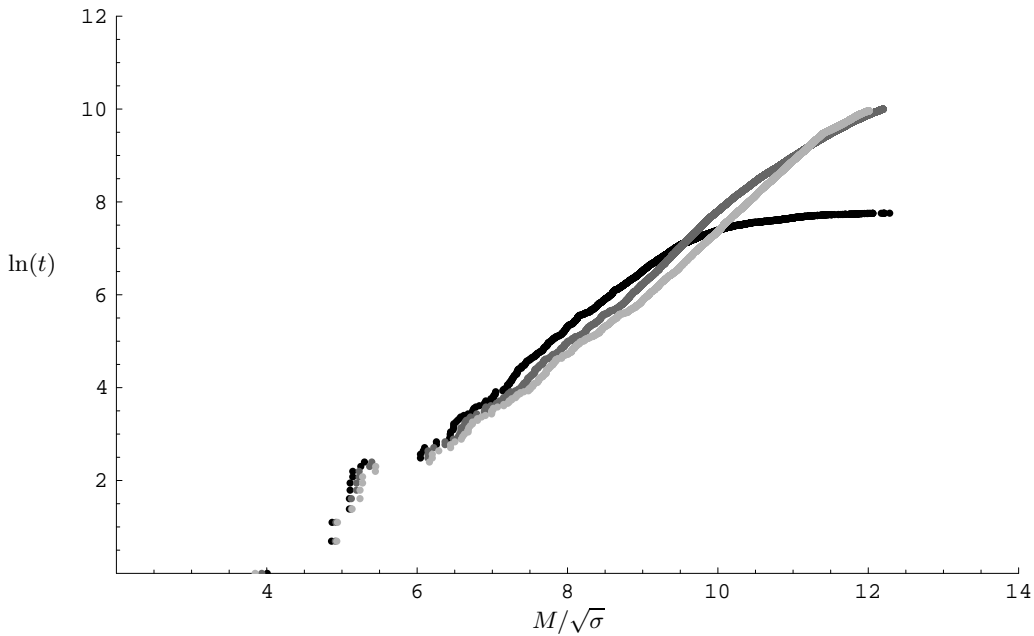}}\\
\hspace{0.5in}$\displaystyle M/\sqrt{\sigma}$
\caption{$\ln(t)$ versus the mass $M_t$ of the $t^{\rm th}$ glueball
in units of the string tension, for $K=8,10,12$ and $a\sqrt{\sigma}=1.44$. 
Lighter data correspond
to larger $K$. 
\label{fig1}}
\end{figure*}

\begin{figure*}
$\displaystyle\ln (t)$\hspace{5pt}
\raisebox{-1.75in}{\includegraphics[width=5in]{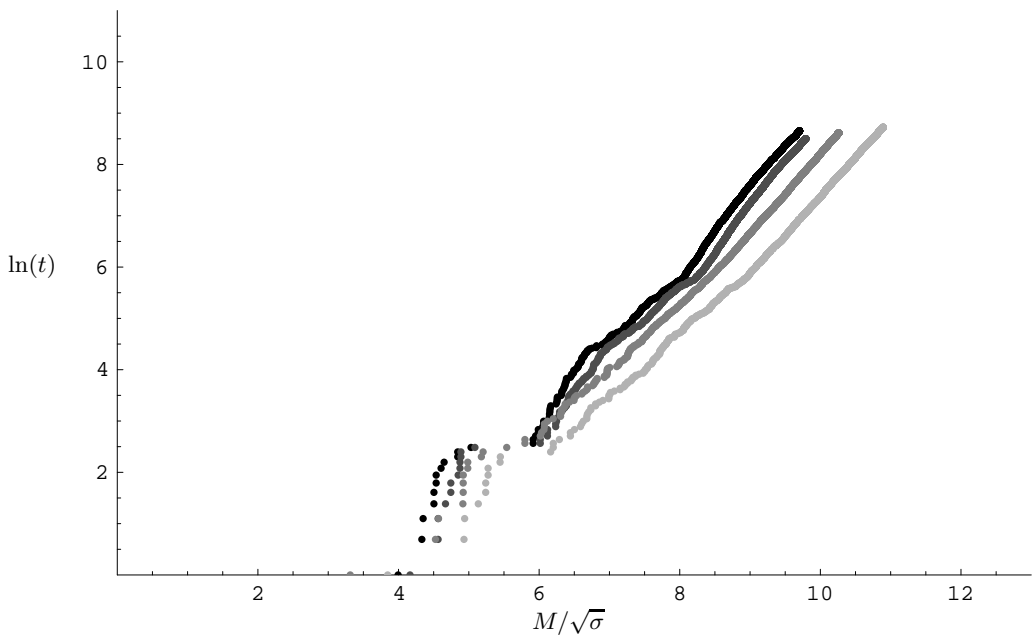}}\\
\hspace{0.5in}$\displaystyle M/\sqrt{\sigma}$
\caption{$\ln(t)$ versus the mass $M_t$ of the $t^{\rm th}$ glueball
in units of the string tension, for $a\sqrt{\sigma} = 1.09,1.21,1.33,1.44$,
and $K=12$. 
Lighter data correspond
to larger $a$.
\label{fig2}}
\end{figure*}

\section{Density of states}

We employ a  DLCQ cutoff $K$ in the $x^-$ direction \cite{dlcq}
to obtain a finite Fock space basis at finite transverse lattice spacing $a$. 
Therefore, 
we must study the stability of  results under variation of both $K$ and
$a$. Figs.~\ref{fig1} and \ref{fig2}
display the distribution of glueball masses $M_t$, in units
of the string tension $\sqrt{\sigma}$, for 
 the $t^{\rm th}$ state in the spectrum. 
The results in Fig.\ref{fig1} are shown for fixed 
$a = 1.44/\sqrt{\sigma}$, the lattice spacing
at which the 
best covariance of the low-lying spectrum was obtained for Lagrangian
(\ref{ferlag}), and increasing
values of the DLCQ cutoff $K=8,10,12$, 
where $K \to \infty$ is the limit of infinite
dimensional Fock space. The largest basis at $K=12$ contains
654,948 states in total, though only the lowest 20,000 are plotted
for file storage reasons.
At the largest masses displayed,
the DLCQ
cutoff is introducing non-universal artifacts because it cuts off
the maximum number of $M$-particles in a state.
At lower masses however, there is universal behavior of the slope of
the distribution.
We repeated the calculation for $K=12$ at four different lattice spacings,
shown in Fig.\ref{fig2},  the couplings of (\ref{ferlag})
re-optimized for covariance of the low-lying spectrum
at each. Again, for the range of masses plotted, the slope is relatively 
universal.

The approximately
universal linear growth observed in this region implies that the density of 
states per unit mass,
$\rho(M) = dt/dM$, grows exponentially with mass $M$
\begin{equation}
\rho(M) \sim {\rm e}^{M/T_c} \ .
\label{exp}
\end{equation}
This exponential behaviour establishes itself already in the first 100
or so states, occuring at less than twice the lightest glueball mass, 
which is within the expected range of validity of the method \cite{dv3}. 
It is perhaps surprising from the figures that the universal
growth appears to extend to much higher masses. 
One would expect the individual highly excited glueball masses to 
shift around
relative to their true values due to cutoff dependence.
However, an aggregate quantity like the density of states appears to be less
sensitive to cutoff dependence.

\section{Critical Temperature and Strings}

It is well known \cite{hagedorn} that exponential growth 
of the density
of states leads to a diverging canonical partition
function for temperature $T>T_c$. 
To determine an accurate value for $T_c$, one needs to know
the density 
for asymptotically high masses $M$, when power corrections
to the expression (\ref{exp}) at lower masses can be neglected. In principle,
knowledge of the functional 
form of power corrections could enable one to extract
$T_c$ accurately from lower masses, but in practice the form is unknown for
QCD. Therefore, although exponential growth is clearly occuring,
we can only extract a rough estimate for $T_c$ from the numerical
data, based on linear
fits to Figs.\ref{fig1} and \ref{fig2}.

There are three systematic sources of error in our estimate. The first
two come from the 
variation of the data in Figs.~\ref{fig1} and \ref{fig2}, which we take
 as indicative
of the error due to residual dependence on the cutoffs $K$ and $a$. 
The third comes from the linear fits themselves, since power corrections 
to Eqn.~(\ref{exp})
in the actual data will make the fit result for $T_c$ depend upon 
the range of masses chosen
for the fit. 
Those intervals are naturally limited above and below: for all the data, 
smooth linear behaviour does not start until the block of points 
beginning at  $M  \approx 6 \sqrt{\sigma}$; 
DLCQ artifacts limit the maximum useful value of $t$, roughly $\ln t=6,7,8$
for the $K=8,10,12$ data respectively. Within these limits, we 
performed  linear fits to the data sets in the figures 
for a variety of different mass intervals. The  maximum 
variation of $T_c$ with  
interval chosen was of order $\pm 15\%$, which we take as indicative of the
error made by using simplified fitting function (\ref{exp}). 
As a result, we estimate $T_c = 0.63 (1)(9)(9)\sqrt{\sigma}$, 
where the first error is due to $K$,
second due to $a$, and third from the fits.

We can also deduce analytic bounds on $T_c$.
In Ref.~\cite{dv2}, we derived in transverse lattice gauge theory 
an approximate analytic formula for the number $d_n$ of highly excited
glueball states of mass squared $M^2 = 2 \pi \sigma n$, $n$ integer.
The derivation was based on the assumption that the glueball masses
were governed only by longitudinal dynamics of the transverse 
link fields and a fixed 
number of such links in a state. 
Assuming a transverse degeneracy factor $g^p$ for the number of 
distinct states with $p$ links, 
the formula of Ref.\cite{dv2}  trivially generalises  to
\begin{equation}
d_n = {\rm exp}\{2\sqrt{-n {\rm Li}_2 [-g^2]}\} \ ,
\label{analytic}
\end{equation}
where ${\rm Li}_2$ is the dilogarithm function.
This leads to a Hagedorn temperature
\begin{equation}
T_c = \sqrt{\pi \sigma \over -2{\rm Li}_2 [-g^2]} \ , 
\label{tc}
\end{equation}
The maximum degeneracy one could expect is from
a long free chain of $p$ links, which on a square
transverse lattice gives $g=4$. This provides an upper bound on the
density of states and
hence a lower bound $T_c = 0.54 \sqrt{\sigma}$. 
Conversely, setting $g=1$ corresponds to neglecting 
transverse structure altogether. 
This is a lower bound on the density of states, implying $T_c =1.375 
\sqrt{\sigma}$ is an upper bound. The numerical solution investigated
above is somewhere between these extremes; in reality there are non-trivial
longitudinal {\em and} transverse dynamics. 

Recent large-$N_c$ extrapolations of
Euclidean lattice computations  \cite{teper} are more suited
to providing accurate determinations of the critical temperature,
obtaining $T_c = 0.596(4) \sqrt{\sigma}$.
This is less than the result, $T_c/\sqrt{\sigma} = \sqrt{3/2\pi} 
\approx 0.69$, for the non-interacting
Nambu-Goto string \cite{alvarez}. 
This would suggest 
 that the large-$N_c$ gauge theory flux string has more worldsheet 
degrees of freedom than a simple
bosonic one, which has only transverse oscillations. 
In making this comparison 
based on $T_c$ with a free string, it is important
to use results from the large-$N_c$ limit, since only in this limit
do lightcone flux strings not split and join. 
At finite $N_c$, by examining only $T_c$
 one cannot separate physically distinct 
world-sheet effects 
from string interaction effects proportional to  $1/N_{c}^2$.
Also, the Hagedorn determination of $T_c$ directly from the density of states
is strictly correct only for a gas of free hadrons/strings.

Although Euclidean lattice results provide 
an accurate value for $T_c$, they do not 
explain why it differs from the bosonic string value. 
Transverse lattice gauge theory can provide  an explanation with the help of
an older
analysis by Klebanov and Susskind \cite{kleb}. They noted that choosing a 
sufficiently tachyonic link field mass $\mu_b$ (see Eqn.\ref{ferlag})
would drive the theory into a new phase, where glueball states 
were unstable to becoming infinitely long chains of transverse links,
each link with frozen longitudinal dynamics. In this phase, there are
only transverse dynamics (this is the complete
opposite of the scenario used above to derive the  bounds). The spectrum
of this phase is exactly that of the Nambu-Goto string \cite{kleb,morris}.
Thus, at least some of  the extra degrees of freedom present
in the QCD phase of transverse lattice gauge theory, which can lower the
critical temperature,  
appear to be longitudinal oscillations of the flux string.

\acknowledgments{
The work of SD is supported by PPARC grant 
PPA/G/0/2002/00470.  The work of BvdS is supported by NSF grant
PHY-0200060.
}

\end{document}